\documentclass[amsmath,amssymb,aps,twocolumn]{revtex4}

\def\be{\begin{equation}}
\def\ee{\end{equation}}
\def\ba{\begin{eqnarray}}
\def\ea{\end{eqnarray}}

\begin{document}

\title{General conditions for scale-invariant perturbations in an expanding universe}

\author{Ghazal Geshnizjani, William H. Kinney,  Azadeh Moradinezhad Dizgah}

\affiliation{Department of Physics, University at Buffalo, SUNY, Buffalo, NY 14260, United States of America}

\date{July 6, 2011}

\begin{abstract}
We investigate the general properties of expanding cosmological models which generate scale-invariant curvature perturbations in the presence of a variable speed of sound. We show that in an expanding universe, generation of a super-Hubble, nearly scale-invariant spectrum of perturbations over a range of wavelengths consistent with observation requires at least one of three conditions: (1) accelerating expansion, (2) a speed of sound faster than the speed of light, or (3) super-Planckian energy density.
\end{abstract}

\preprint{}



\maketitle

\section{Introduction}
Observations of both the Cosmic Microwave Background (CMB) and large-scale structure (LSS) are compatible with a spectrum of nearly scale-invariant and Gaussian primordial perturbations with correlations on super-Hubble length scales \cite{Komatsu:2010fb}, in agreement with the predictions of the simplest inflationary models. However, inflation is not the only way to generate cosmological perturbations. Alternatives that have been put forward range from contracting scenarios where the universe needs to go through a singular or non-singular bounce (see \cite{Khoury:2011ii} and references there), string gas cosmology, in which the universe is initially static \cite{Nayeri:2005ck}, varying fundamental speed of light \cite{Albrecht:1998ir}, or a rapidly varying speed of sound \cite{ArmendarizPicon:2003ht,ArmendarizPicon:2006if,Magueijo:2008pm, Bessada:2009ns, Magueijo:2009zp}. All of the alternatives so far suggest that to produce the observed primordial power spectrum, any route but inflation requires an understanding of gravity beyond general relativity. This is either due to violation of the Null Energy Condition, a singular bounce, or the breaking of Lorentz invariance due to a preferred frame of reference to explain  varying speed of light or super-luminal sound speed \cite{Babichev:2007dw}. (It is worth noting that inflation itself, while more successful than its alternatives on many fronts, still has its own challenges to overcome, ranging from setting the correct initial conditions \cite{Vachaspati:1998dy} to relying on the existence of a scalar field which has not yet been confirmed by any fundamental theory.) It is interesting to ask: what is the most general conclusion that can be drawn from the observed spectrum of primordial perturbations in the universe?

In this paper, we study general properties of the production of a scale-invariant two point function. We use the simple framework  put forward in Ref. \cite{Khoury:2008wj} and later used in Ref. \cite{Khoury:2010gw} that makes it simple to study scenarios  that have different background evolution but result in the same second-order gravitational action for quantum perturbations. The format of the paper is the following: In Section \ref{setup} we review the framework developed in Ref. \cite{Khoury:2008wj}. In Section \ref{nogo} we prove that to produce enough scale invariant modes  on super-Hubble scales in an expanding universe one of the following conditions must be met: (1) accelerating expansion, (2) superluminal sound speed, or (3) super-Planckian energy density. Section \ref{conclusion} contains discussion and conclusions.


\section{  Action and solutions for scale-invariant curvature perturbations}  \label{setup}
The quadratic action for curvature perturbation $\zeta$ around a flat Friedmann-Robertson-Walker (FRW) background with a time dependent sound speed $c_s(\tau)$ in general can be written as \cite{Brandenberger:1993zc, Garriga:1999vw} :
\be \label{action1}
S_2=\frac{M_{pl}^2}{2}\int dx^3 d\tau~ z^2 \left [ \left (\frac{d\zeta}{d\tau}\right)^2- c_s(\tau)^2(\nabla\zeta)^2 \right ], 
\ee
where
\be
 z\equiv \frac{a\sqrt{2 \epsilon}}{c_s}, \label{z}.
\ee
Here $a$ and $\tau$ are the scale factor and conformal time,  respectively, while $\epsilon$ is defined through Hubble  parameter $\epsilon\equiv {-\dot{H}/H^2}$.  

Note that for $\epsilon < 0$, which corresponds to phantom matter, we will have $z^2<0$.   The overall sign of the action does not change the equation of motion and it leads to the same two-point function for $\zeta$. However a problem arises, since in the presence of other fields, the kinetic term of $\zeta$ has an opposite sign. By quantizing those, we will end up with Hamiltonians which are unbounded in opposite directions, with ghost instabilities. 

As was pointed out in \cite{Khoury:2008wj}, through a time transformation $dy=c_s d\tau$ one can re-express the action (\ref{action1}) in the new form:

\be
S_2={M_{pl}^2\over 2}\int dx^3 dy~ q^2 \left [ \left ({d\zeta\over dy}\right)^2- (\nabla\zeta)^2 \right ], 
\ee
where  
\ba\label{q}
q\equiv {a\sqrt{2 \epsilon}\over \sqrt{c_s}}.
\ea 

Now introducing the canonically normalized scalar variable $v= M_{pl}~ q ~\zeta$, the equation of motion for $v$ in Fourier space is given by 
\be
v_k''+(k^2-{q''\over q})v_k=0, 
\ee
where prime represents $d/dy$. 

A scale-invariant spectrum is obtained if 
\be\label{qequation}
{q''\over q}\sim {2\over y^2},
\ee
 in analogy with  canonical inflation where $z_{,\tau\tau}/z\sim 2/\tau^2$. Eq. (\ref{qequation}) has a general solution of the form  
\be \label{solq}
q={\alpha\over y}+\beta y^2,
\ee for arbitrary $
\alpha$ and $\beta$. 

Production of superhorizon perturbations requires the following conditions: At early time, a mode with given comoving wavenumber $k$ starts in an approximately Minkowski vacuum state with $k^2 \gg 2 / y^2$, such that the WKB approximation is valid. At late time, the mode evolves such that $k^2 \ll 2 / y^2$ and WKB breaks down, resulting in particle production.  Since for increasing time $d \tau > 0$, $dy > 0$,  we see that  for particle production we must have  $y \in [-\infty , 0]$. 
Therefore, at early times $\omega_k^2\sim k^2$ and  is almost constant,   so that $|\omega_k'/\omega_k^2| \ll 1$ and WKB is correct. However,  at late times $\omega_k^2\sim -2/y^2$, $|\omega_k'/\omega_k^2|\sim {\cal O}(1)$ and  WKB is no longer valid.  This argument already enables us to see that    at late time the $\alpha/y$ term is always an attractor solution  and dominates over $\beta y^2$. 

Applying the standard method of quantization for $v$ and setting the adiabatic initial condition at early times, the familiar mode function \cite{Khoury:2008wj} is obtained:
\be
v_k(y)={1\over \sqrt{2k}}\left (1-{i\over ky}\right)e^{-iky}. 
\ee
In long wavelength limit $v_k$ behaves as 
\be \label{nuIRlimit}
v_k\sim {-1\over \sqrt{2k}}\left ({i\over ky}+{1\over 2}k^2y^2\right)~~~~ ky\ll1.
\ee
Note that mode freezing occurs when $k y = 1$, which is not the same as when the mode crosses the Hubble horizon, $k = a H$. There are {\it three} horizons with independent dynamics: the Hubble horizon, $R_H \equiv 1/ (a H)$, the sound horizon, $R_s\equiv c_s / (a H)$, 
and the ``freezeout horizon''\footnote{We use the terminology of freezeout horizon loosely to be consistent with literature. However, note that in the case where $\beta\neq0$, then $\zeta$ is not conserved.} 
\be
R_\zeta\equiv - y.
\ee
All three horizons are expressed in comoving units. As long as the freezeout horizon is shrinking, generation of perturbations occurs. Furthermore, modes can be generated on super-Hubble scales even if the Hubble horizon is growing, as long as the freezeout horizon is larger than the Hubble horizon \footnote{For an example of an expanding scenario without super luminal speed of sound where the freezeout horizon is larger than the Hubble horizon see \cite{Khoury:2010gw}.}.
We can now calculate the power spectrum ${\cal P}^\zeta_k\equiv k^3\zeta_k^2 $ for different choices of $\alpha$ and $\beta$.
First, taking $\beta=0$ is the case very similar to inflation since for superhorizon modes we obtain:
\be
\label{powerspalpha}
k^3\zeta_k^2\sim k^3\left( {\nu_k \over M_p q}\right)^2 \sim {1\over 2M_p^2\alpha^2},
\ee
which is the exact scale-invariant power spectrum. The   amplitude of $\zeta$ is constant outside freezeout horizon,  $R_\zeta\equiv \sqrt{q/q^{\prime\prime}}\sim y$.  This solution leads to a stable solution for the background spacetime, since at $k \rightarrow 0$, a constant metric perturbation does not change the time evolution of the scale factor. 

Next we consider $\alpha=0$. Substituting for the power spectrum leads to:
\be
k^3\zeta_k^2\sim k^3\left( {\nu_k \over M_p q}\right)^2 \sim {1\over M_p^2\beta^2 y^6}. 
\ee
While this spectrum is scale invariant, the amplitude of $\zeta$ grows outside $R_\zeta$ which signals instability or  a non-attractor behavior for  the background.  

Last, if neither of $\alpha$ or $\beta$ are zero then 
\be
k^3\zeta_k^2\sim k^3\left( {\nu_k \over q}\right)^2 \sim {1\over 2M_p^2}\left ({1\over \alpha +\beta y^3}\right)^2. 
\ee
At late times, $y^3\ll \alpha/\beta$, the $\alpha/y$ term will win over $\beta y^2$, and we regain the conserved $\zeta$ solution. Since we also have $ky\ll1$ this condition is automatically satisfied if $k^3>\beta/\alpha$. So   in this case    the amplitude is also well behaved and not divergent as $y\rightarrow 0$. 

It is also worth noting that all these solutions correspond to exact scale invariance. In other words , even though applying the naive slow-roll result $n_s-1\sim 2(\epsilon+d \ln{\epsilon}/Hdt+d \ln{c_s}/Hdt )$ may suggest    otherwise, calculating  the spectral index $n_s$ would result in 
\be
n_s-1=0.
\ee
In practice, to obtain non-zero tilt, $q$ must have deviations from these solutions. For example taking $\beta=0$ and allowing for $\alpha$ to have small time dependence can lead to a small tilt \cite{Hu:2011vr}. 


\section{General conditions for scale invariance} \label{nogo}
 
This section contains the main result of the paper: In an expanding universe, in order to generate a scale-invariant spectrum of curvature perturbations on a range of scales compatible with observations, one of three conditions must be met: (1) accelerating expansion ({\it i.e.} inflation), (2) speed of sound faster than the speed of light, or (3) super-Planckian energy density. Current observations of CMB and LSS indicate that the spectrum of curvature perturbations must be nearly scale-invariant over {\it at least} three decades in wavelength, and we will take this to be the lower bound.

   We first consider the simple case of $c_s = 1$ in an expanding background. Non-accelerating expansion implies that Hubble  horizon measured in comoving units  is always growing, since $\epsilon>1$:
\be \label{R_H}
{d R_{H}\over d\tau}=\epsilon -1 > 0.
\ee 
Following the framework described in Sec. \ref{setup}, production of scale invariant modes for  $c_s=1$ leads to  the freezeout horizon shrinking as $R_\zeta\equiv \sqrt{z/z_{\tau\tau}}\sim - \tau$. Note that the same argument that we used in that section for the range of $y$  applies here  to the range of conformal time $\tau$ such that   $\tau \in [-\infty , 0]$. 
 
Consider modes with comoving wavelengths $\lambda_f<\lambda < \lambda_i$ corresponding to the scale-invariant modes that we observe today in CMB and LSS.  Observations of CMB and LSS    require scale invariance over about three decades in wavelength,
\be
\label{maxLambda}
 \lambda_{i} \gtrsim 1000 ~\lambda_f.  
 \ee
Since the horizon is always growing, for the modes to be larger than the Hubble horizon at late time, they must also be superhorizon at early time, so that 
\be
\label{superHubble}
\lambda_f\left(\tau_f\right) > R_H\left(\tau_f\right).
\ee
Here the conformal time $\tau$ is taken to be the time when a given mode crosses the freezeout horizon, $\lambda_i \sim |\tau_i|$, $\lambda_f \sim |\tau_f|$. Then the conditions (\ref{maxLambda}) and (\ref{superHubble}) become:
 \ba 
 |\tau_{i}|&\gtrsim& 1000 ~|\tau_f |, \\ |\tau_f |&>&  R_H(\tau_f),
 \ea
 which implies 
 \be 
\label{taurange}
 {\tau_f-\tau_i\over R_H(\tau_f)}> 1000.
 \ee
 Now, writing  the continuity equation
 \be
 {\dot{\rho}\over \rho}=-2\epsilon H,
 \ee
 integrating from $\tau_i$ to $\tau_f$ leads to the   inequality
\ba \label{densitytime}
\ln{{\rho_i\over\rho_f}}&=&2\int_{t_i}^{t_f} \epsilon H dt \nonumber\\
 &=&2\int_{\tau_i}^{\tau_f}{ \epsilon R^{-1}_H(\tau) ~ d\tau} \nonumber\\
 &>& 2 R^{-1}_H(\tau_f)\int_{\tau_i}^{\tau_f}{\epsilon ~ d\tau} \nonumber\\
 &>& 2 R^{-1}_H(\tau_f)\epsilon_{\rm min} (\tau_f-\tau_i),
\ea
where $\epsilon_{\rm min}$ is the minimum value of $\epsilon$, and we have used the fact that non-accelerating expansion requires $R_H(\tau) < R_H(\tau_f)$. Since non-accelerating expansion also means that $\epsilon_{\min} \geq 1$, the relation (\ref{taurange}) results in the following inequality for the cosmological density:
\be
\ln{{\rho_i\over\rho_f}} > 2000,
\ee
or
\be
\rho_i > 10^{868} \rho_f.
\ee 
Taking the lower bound on $\rho_f$ to be given by the lower bound on the reheat temperature, which is given by Big Bang nucleosynthesis, $\rho_f > \rho_r > (100\ {\rm MeV})^4$, we have 
\be
\rho_i \gg M_{\rm Pl}^4.
\ee
This is purely the result of our two assumptions, non-accelerating expansion, and mode generation on a sufficiently large range of scales (\ref{maxLambda}).  Therefore, we have shown that to produce enough super-Hubble modes at reheating,  the initial density of our scenario has to start larger than the Planck energy for decelerating expansion ($\epsilon_{\rm min} \geq 1$). For larger values of $\epsilon$, the problem becomes more severe.  Furthermore, if we want the    range of modes (\ref{maxLambda}) to be one order of magnitude larger we need $e^{10}$    higher energy density $\rho_i$. It is also interesting that this problem can arise in a different form in contracting scenarios as well:    even though the density is sub-Planckian the curvature  still    becomes exponentially greater than the Planck curvature \cite{Linde:2009mc}.

Next, we repeat the same calculation allowing $c_s$ to vary.  The modes this time exit the freezeout horizon when  $\lambda \sim |y|$. 
Therefore, the conditions (\ref{maxLambda}) and (\ref{superHubble}) now yield: 
  \ba 
 |y_{i}|&\gtrsim& 1000 ~|y_f | \\ |y_f |&>&  R_H(\tau_f),
 \ea
 which implies 
 \be 
\label{YR}
 {y_f-y_i\over R_H(\tau_f)}> 1000.
 \ee
The inequality (\ref{densitytime}) is still valid and since 
\be
y_f-y_i=\int_{\tau_i}^{\tau_f} c_s d\tau =  {\bar c_s}(\tau_f-\tau_i), 
\ee
where ${\bar c_s}$ is the average sound speed, we obtain:
\be
\ln{{\rho_i\over\rho_f}} > 2 R^{-1}_H(\tau_f)\epsilon_{\rm min} \frac{y_f-y_i}{\bar c_s}.
\ee
Combining above condition with (\ref{YR}), $\epsilon_{\min} >1$,  we have  
\be
 \frac{2000}{\bar c_s} <\ln{{\rho_i\over\rho_f}} < \ln{\frac{M_{Pl}^4}{\rho_r}} \sim 80 \ln{10},
\ee
where we take $\rho_r \sim (100\ {\rm MeV})^4$. This results in a lower bound on the average sound speed,
\be
{\bar c_s} > 10. 
\ee
   Therefore, we have shown that in  a non-accelerating expanding universe, if the energy density starts sub-Planckian, producing the range of scale invariant modes consistent with observations requires super-luminal sound speed.

Could a small deviation from scale invariance as favored by the WMAP 7-year data \cite{Komatsu:2010fb} weaken these bounds? For a power-law spectrum $P(k) \propto k^{n - 1}$, the freezeout horizon behaves as 
\be
R_\zeta^{-2} = \frac{q''}{q} = \frac{2 + (3/2) \left(1 - n\right)}{y^2} \propto \frac{1}{y^2},
\ee
so that the bound remains essentially unchanged even in the case of weakly broken scale invariance.  The next section presents discussion and conclusions. 

\section{Concluding remarks} 
\label{conclusion} 

The universe is observed to have a spectrum of nearly scale-invariant density perturbations over about three decades in wavelength,   
 which were at scales larger than the Hubble length  at early times. In this paper, we have shown that generating cosmological perturbations consistent with observation in an expanding universe requires at least one of: (1) accelerated expansion, (2) superluminal sound speed, or (3) super-Planckian energy density. We note that this applies to the curvaton mechanism \cite{Lyth:2001nq} as well, since the freezeout horizon for a free scalar ``spectator'' field shrinks as $\tau$ ($a \propto 1/\tau)$ to produce scale-invariant perturbations. Therefore super-Hubble curvaton fluctuations directly require inflation.
It is important to note that scale invariance alone does not require inflation or ``tachyacoustic'' \cite{Bessada:2009ns} evolution: a key point is that even with sub-luminal sound speed, super-Hubble perturbations can be generated with a growing Hubble horizon, as long as the freezeout horizon is shrinking and $R_\zeta > R_H$ \cite{Khoury:2010gw}. However, one cannot generate three decades of modes in this fashion without super-Planckian energy densities. Perturbations could also be generated on sub-Hubble scales by a period of non-inflationary expansion \cite{Khoury:2010gw,Khoury:2011ii}, and only later redshifted to super-Hubble scales by a subsequent period of inflation. This would be consistent with our bound. Recent work has, however, shown why seeding  scale-invariant fluctuations with sub-luminal sound speed is challenged by the breakdown of weak coupling and therefore perturbation theory \cite{Baumann:2011dt}. This issue is quite generic and can arise even in non-expanding scenarios (see \cite{Khoury:2011ii}). Similarly, string gas cosmology requires accelerating expansion to exit from the initial, static phase, but perturbations are not generated by the usual inflationary mechanism.  Our bound also does not apply to contracting cosmologies. However, in the case of an expanding cosmology, we have presented a simple and general argument showing why either a superluminal sound speed or a period of inflation is required for the successful generation of a scale-invariant, super-Hubble spectrum of perturbations.


\section*{Acknowledgements}
We thank Niayesh Afshordi, Justin Khoury and Sara Shandera for useful discussions. This research is supported  in part by the National Science Foundation under grant NSF-PHY-0757693.

\bibliography{SCINC-1}

\end{document}